\documentstyle[prb,aps,epsf,floats]{revtex} \def\narrowtext{} \tighten 
\twocolumn
\input epsf.sty
   
\begin{document}
\draft

\title{Shadow bands in single-layered Bi$_2$Sr$_2$CuO$_{6+\delta}$  studied by angle-resolved photoemission spectroscopy}

\author{
	K. Nakayama,$^1$
	T. Sato,$^{1,2}$
	T. Dobashi,$^1$
	K. Terashima,$^1$
	S. Souma,$^{1,3}$
	H. Matsui,$^1$
	T. Takahashi,$^{1,2}$\\
	J. C. Campuzano,$^3$
	K. Kudo,$^4$
	T. Sasaki,$^4$
	N. Kobayashi,$^4$
	T. Kondo,$^5$
	T. Takeuchi,$^{5,6}$\\
	K. Kadowaki,$^7$
	M. Kofu,$^8$
	and K. Hirota$^9$
	}
\address{
	$^1$Department of Physics, Tohoku University, Sendai 980-8578, Japan\\
	$^2$CREST, Japan Science and Technology Agency (JST), Kawaguchi 332-0012, Japan\\
	$^3$Department of Physics, University of Illinois at Chicago, Chicago, Illinois 60607, USA\\
	$^4$Institute for Materials Research, Tohoku University, Sendai 980-8577, Japan\\
	$^5$EcoTopia Science Institute, Nagoya University, Nagoya 464-8603, Japan\\
	$^6$Department of Crystalline Materials Science, Nagoya University, Nagoya 464-8603, Japan\\
	$^7$Institute of Materials Science, University of Tsukuba, Ibaraki 305-3573, Japan\\
	$^8$Institute of Materials Structure Science, KEK, Tsukuba 305-0801, Japan\\
	$^9$Institute for Solid State Physics, University of Tokyo, Kashiwa, Chiba 277-8581, Japan\\
	}

\address{
\begin{minipage}[t]{6.0in}
\begin{abstract}
We have performed systematic angle-resolved photoemission spectroscopy (ARPES) on single-layered cuprate superconductor Bi$_2$Sr$_2$CuO$_{6+\delta}$ to elucidate the origin of shadow band.  We found that the shadow band is exactly the $c$(2$\times$2) replica of the main band irrespective of the carrier concentration and its intensity is invariable with respect to temperature, doping, and substitution constituents of block layers.  This result rules out the possibility of antiferromagnetic correlation and supports the structural origin of shadow band.  ARPES experiments on optimally doped La$_{1.85}$Sr$_{0.15}$CuO$_4$ also clarified the existence of the $c$(2$\times$2) shadow band, demonstrating that the shadow band is not a unique feature of Bi-based cuprates.  We conclude that the shadow band is related to the orthorhombic distortion at the crystal surface.
\typeout{polish abstract}
\end{abstract}
\end{minipage}}
\maketitle
\narrowtext

\begin{figure*}[!t]
\epsfxsize=6.8in
\epsfbox{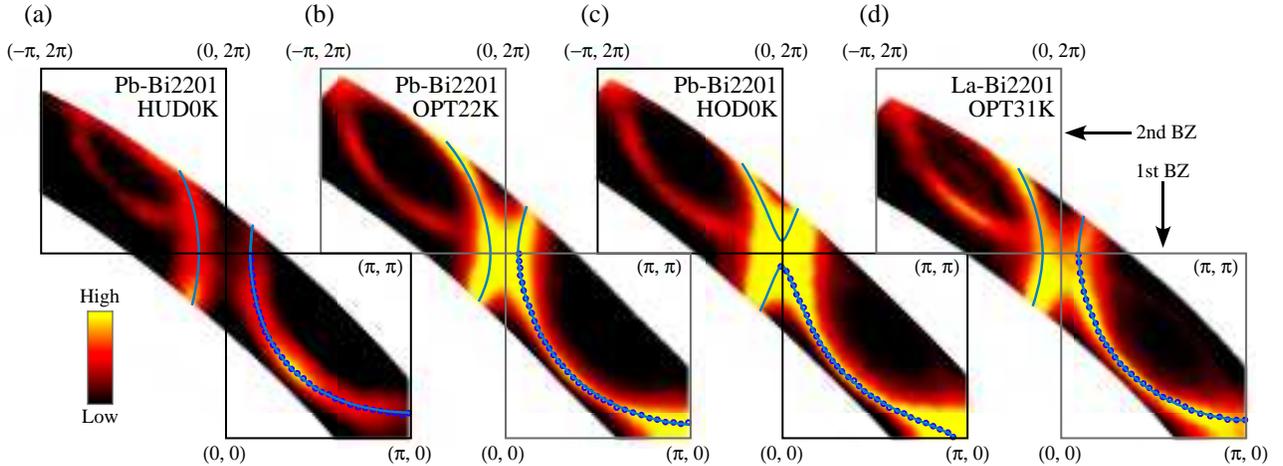}
\vspace{0.25cm}
\caption{
ARPES spectral intensity plots as a function of two-dimensional wave vector for (a) heavily underdoped Pb-Bi2201 (HUD0K), (b) optimally doped Pb-Bi2201 (OPT22K), (c) heavily overdoped Pb-Bi2201 (HOD0K), and (d) optimally doped La-Bi2201 (OPT31K).  The ARPES spectrum is normalized by the spectral intensity at high energy (400 meV) as in previous study,\cite{Kordyuk} and the intensity at $E_{\rm F}$ is obtained by integrating the spectral intensity over the energy range of 20 meV with respect to $E_{\rm F}$.  Small circles represent estimated $k_{\rm F}$ positions of the main band in the first BZ.  Solid curves show FSs determined by fitting the experimentally-obtained $k_{\rm F}$ positions with a tight-binding model. 
}
\label{Fig. 1(Color)}
\end{figure*}

\section{INTRODUCTION}
Fermi surface (FS) topologies are directly connected to various physical properties such as transport and thermodynamics, and further give an experimental base for better understanding anomalous physical properties and developing theoretical models to explain the exotic phenomena.  In high-$T_c$ cuprate superconductors (HTSCs), the topology of FS has been intensively studied by high-resolution angle-resolved photoemission spectroscopy (ARPES), and such investigation has been mainly focused on Bi$_2$Sr$_2$CaCu$_2$O$_{8+\delta}$ (Bi2212) which contains double CuO$_2$ layers in a unit cell.  It has been revealed that FS of Bi2212 is characterized by three different components, a large holelike FS centered at ($\pi$, $\pi$), its diffraction replica by a superstructure of BiO layers referred as umklapp FSs, and a shadow FS which is a replica of the main FS transferred by $Q$ = ($\pi$, $\pi$).\cite{Aebi}  Recent progress in energy and momentum resolutions of ARPES further enables us to directly observe the bilayer splitting due to the interaction between two CuO$_2$ planes \cite{Feng,Chuang} and the transition of FS topology in the heavily overdoped region.\cite{Kaminski}  In spite of these remarkable experimental progresses, there are still fierce debates on the origin of shadow FS or shadow bands.  It has been proposed that the short-range antiferromagnetic (AF) correlation in CuO$_2$ layers is responsible for the emergence of shadow band owing to the $Q$=($\pi$, $\pi$) nature of the shadow band.\cite{Aebi,Kampf,Langer,Chubukov,Stephan}  Recently, objections against the AF scenario from the structural point of view have been reported.\cite{Chakravarty,Vilk,Singh}  However, previous experimental results and their interpretations are still highly controversial.  Kordyuk $et$ $al$.\cite{Kordyuk} reported by doping-dependent ARPES of Bi2212 that the strength of shadow band mirrors $T_c$ values, suggesting the close link between the shadow band and superconductivity, whereas there are conflicting reports showing that the shadow band intensity is doping-independent.\cite{Schwaller,Izquierdo}  LaRosa $et$ $al$. reported that the intensity of shadow band depends on the binding energy in favor of the AF picture,\cite{LaRosa} while Koitzsch $et$ $al$. reported that it is invariant with respect to the binding energy.\cite{Koitzsch}  As for the origin of the structural distortions, two essentially different explanations, a final-state effect due to photoelectron diffraction at the surface \cite{Chakravarty,Izquierdo} and an initial-state effect due to orthorhombic distortion,\cite{Koitzsch,Mans} have been proposed to date.  A recent ARPES study of Bi2212 by Mans $et$ $al$. with tunable light polarization \cite{Mans} concluded that the shadow band is due to the orthorhombic distortion from the tetragonal symmetry.  This result certainly opens a way to better understanding the nature of shadow band in Bi2212.  However, there are still unresolved problems as to (i) the possible link between the character of shadow band and important physical parameters such as $T_c$ or superconducting-gap magnitude, (ii) whether the shadow band is created by a single mechanism independent of doping concentration, and (iii) whether the emergence of shadow band is a universal feature of all HTSCs.  In order to answer these essential questions, it may be important to investigate the shadow band of HTSC different from intensively studied Bi2212.  In this sense, single-layered (Bi,Pb)$_2$Sr$_2$CuO$_{6+\delta}$ (Bi2201) has several advantages; (i) it is free from the contamination of electronic states by the bilayer splitting and the umklapp bands, (ii) the low-temperature normal state is available due to the low maximum $T_c$ ($\sim$ 30 K), and (iii) the accessible doping region is wide enough to cover the whole superconducting dome.
	 
	In this paper, we report systematic ARPES results as a function of doping and temperature on superstructure-free single-layered Pb-substituted Bi2201 (Pb-Bi2201) and La-substituted Bi2201 (La-Bi2201).  We found that the pronounced shadow band clearly exists even in the nonsuperconducting heavily overdoped region where the AF correlation is significantly weak.  The strength of shadow band is independent of doping range and the temperature, although it suffers a strong $k$-dependent matrix-element effect.  We also found an evidence for the existence of shadow band in optimally doped La$_{1.85}$Sr$_{0.15}$CuO$_4$ (LSCO), indicating that the shadow band is not a unique feature of Bi-based HTSCs.

\section{EXPERIMENTS}
	High-quality single crystals of superstructure-free (Bi,Pb)$_2$Sr$_2$CuO$_{6+\delta}$ (Pb-Bi2201) and optimally doped (Bi,Pb)$_2$(Sr,La)$_2$CuO$_{6+\delta}$ ($T_c$ = 31 K; La-Bi2201) were grown by the floating-zone method.\cite{Kudo,KondoQP,KondoFS}  The hole concentration was controlled by annealing the samples under vacuum or oxygen atmosphere at high temperature.  The $T_c$ of samples was determined by the magnetic susceptibility measurement.  Samples are labeled by their doping levels (HUD for heavily underdoped, OPT for optimally doped, and HOD for heavily overdoped), together with their onset $T_c$.  For example, OPT22K means an optimally doped sample with the $T_c$ of 22 K.  0 K in HUD0K and HOD0K samples means that the sample does not show any signature of superconductivity down to 1.5 K.  Details of sample preparation method for Bi2212 and LSCO have been described elsewhere.\cite{Kadowaki,Kofu}
	
	ARPES measurements of Bi2201 and Bi2212 were performed using a GAMMADATA-SCIENTA SES-200 spectrometer with a high-flux discharge lamp and a toroidal grating monochromator.  We used the He-I$\alpha$ ($h\nu$ = 21.218 eV) resonance line to excite photoelectrons. ARPES measurements of LSCO have been done using a GAMMADATA-SCIENTA R4000 spectrometer at PGM (plane grating monochromator) beamline in synchrotron radiation center, Wisconsin.  The energy and angular resolution were set at 20 meV and 0.2$^\circ$-0.3$^\circ$, respectively.  Sample orientations were determined by the Laue x-ray-diffraction pattern prior to the ARPES measurement.  We did not find any superlattice diffraction spots in Pb-Bi2201 samples, confirming the absence of 5$\times$1 superstructure in BiO layers.\cite{Yamamoto}  Clean surfaces for ARPES measurements were obtained by $in$ $situ$ cleaving of crystals in an ultrahigh vacuum better than 6.5$\times$10$^{-11}$ Torr.  The Fermi level ($E_{\rm F}$) of the sample was referenced to that of a gold film evaporated onto the sample substrate.

\section{RESULTS}

	Figure 1 shows ARPES spectral intensity plots at $E_{\rm F}$ as a function of two-dimensional wave vector for Pb-Bi2201 with three different doping levels and La-Bi2201, measured with the He I$\alpha$ resonance line ($h\nu$ = 21.218 eV).  Experimental data have been obtained over a wide-$k$ region from the first Brillouin zone (BZ) to the second BZ.  As seen in Figs. 1(a) and 1(b), a large holelike main FS centered at the ($\pi$, $\pi$) point is observed in the HUD0K and OPT22K samples of Pb-Bi2201.  A clear signature of shadow FS is found in the second BZ, although its intensity is markedly suppressed in the first BZ due to the matrix-element effect.\cite{Bansil}   We do not observe multiple FSs produced by the umklapp process of photoelectrons through the BiO layer, confirming the absence of 5$\times$1 superlattice modulation.\cite{Yamamoto}  As seen in the HUD0K sample [Fig. 1(a)], the intensity around the antinodal region is weaker than that at the nodal region, reflecting the opening of a pseudogap.  As the hole concentration is increased, the intensity around ($\pi$, 0) gradually recovers [Figs. 1(b) and 1(c)] due to the closure of the pseudogap and the emergence of a sharp quasiparticle peak.\cite{KondoQP,YoshidaQP,Zhou}  In order to determine the exact location of the FS, we have estimated $k_{\rm F}$ positions of the main band by fitting the momentum distribution curves (MDC) at $E_{\rm F}$, except those for the antinodal region of HOD samples which were determined by tracing the peak position of energy distribution curves (EDCs) divided with the Fermi-Dirac (FD) function, \cite{KondoFS,Satotemp,Takeuchi2D3D} since the MDC analysis is not applicable because the energy position of the main band is very close to $E_{\rm F}$.\cite{Satotemp}  The estimated $k_{\rm F}$ positions are shown by circles in Fig. 1.  Next we have fit these $k_{\rm F}$ points by tight-binding formula,\cite{Norman} as indicated by solid curves in Fig. 1.  As seen in Figs. 1(a)-1(c), the volume of the hole FS systematically expands and the topology shows a distinct transition from holelike to electronlike, consistent with recent ARPES reports on Bi2201,\cite{KondoFS,Takeuchi2D3D} and other HTSCs.\cite{Kaminski,Ino,YoshidaFS}  It is noted that appearance of shadow FS is robust to doping-dependent changes of main FSs.  The shadow band exists in a wide range of doping level from heavily underdoped to heavily overdoped region where superconductivity actually vanishes, demonstrating that there is no direct relation between the existence of shadow band and the occurrence of superconductivity.  As seen in Fig. 1(d), we observed a similar shadow band in La-Bi2201 where Sr ions in SrO layers are partially (20$\%$) substituted with La.  This fact also suggests that the appearance or absence of shadow FS is insensitive to the substitution constituents of block layers.

\begin{figure}[!t]
\begin{center}
\epsfxsize=2.6in
\epsfbox{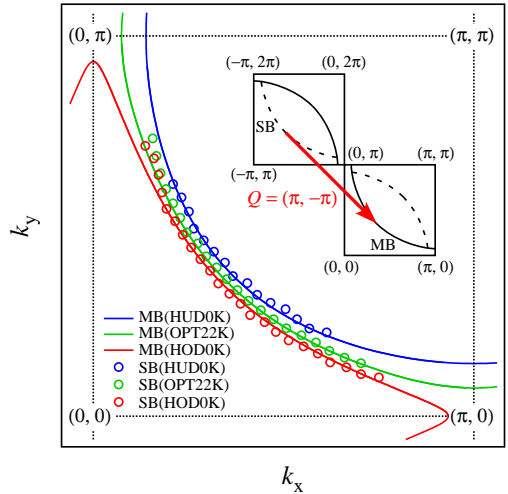}
\end{center}
\vspace{0.25cm}
\caption{
Comparison between $k_{\rm F}$ points of shadow band (SB; open circles) and the experimentally-obtained FS of main band (MB; solid curves) for different doping levels.  $k_{\rm F}$ points of the shadow band in the second BZ are shifted by $Q$ = ($\pi$, $-\pi$) as indicated by the inset.  FS of the main band is obtained by the tight-binding fitting of $k_{\rm F}$ points of the main band as shown in Fig. 1.  The size of circles show the experimental uncertainty in determining the location of $k_{\rm F}$ in the shadow band.
}
\label{Fig. 2(Color)}
\end{figure}

	In order to clarify the relation between the main and shadow FSs, we compare in Fig. 2 the Fermi vectors between the main band and shadow band.  $k_{\rm F}$ points of shadow FS (open circles) are shifted by $Q$ = ($\pi$, $-\pi$) which corresponds to a real-space  periodicity of $c$(2$\times$2), and superimposed on the experimentally determined main FS (solid curves; the tight-binding fittings of $k_{\rm F}$ points of the main band in Fig. 1).  Absence of experimental $k_{\rm F}$ points for the shadow FS around the antinodal region is due to the difficulty in identifying them because of overlapping from the main FS.  As seen in Fig. 2, it is clear that the shadow FS coincides well with the main FS within the experimental uncertainty irrespective of the hole concentration.  This demonstrates that the shape of shadow FS indeed reflects a doping-dependent evolution which fairly tracks the location of the main FS.  These experimental results unambiguously indicate that, at least away from the ($\pi$, 0) point, the shadow FS is a $c$(2$\times$2) replica of the main FS.

\begin{figure}[!t]
\epsfxsize=3.4in
\epsfbox{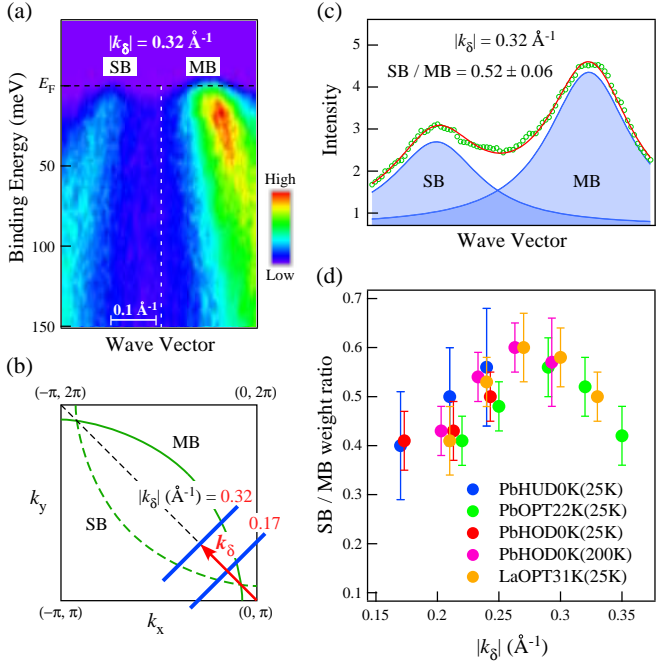}
\vspace{0.25cm}
\caption{
(a) ARPES intensity plot of Pb-Bi2201 (OPT22K) at 25 K as a function of binding energy and wave vector, measured along a cut with $\left| k_{\delta}\right|$ = 0.32 ${\rm \AA}^{-1}$.
(b) Schematic view of the main and shadow FSs in a second BZ, together with the definition of $k_{\delta}$ (red arrow).  Spectra were measured along momentum cuts (blue lines) perpendicular to the direction of $k_{\delta}$.  The black dashed line shows the $c$(2$\times$2) zone boundary.
(c) MDC at $E_{\rm F}$ derived from experimental data in Fig. 3(a) (green open circles).  Red curve is a result of fitting of MDC with two Lorentzians (filled blue curves) with a constant background.
(d) Plot of the spectral-weight ratio of the shadow band to the main band as a function of $\left| k_{\delta}\right|$, for Pb-Bi2201 and La-Bi2201, measured at various conditions with varying doping, temperature, and substitution constituents of block layers (Pb or La).  The spectral weight of each band is obtained by integrating each Lorentzian peak.  Note that estimation of the ratio around the antinodal region is difficult because of the proximity of shadow and main FSs.
}
\label{Fig. 3(Color)}
\end{figure}

     To discuss the character of the shadow FS more quantitatively, we performed numerical analyses on the intensity ratio between the shadow band and main band.  Figure 3(a) shows the ARPES intensity plot as a function of binding energy and wave vector for Pb-Bi2201 (OPT22K) at 25 K, measured at a cut in the second BZ with $\left| k_{\delta}\right|$ = 0.32 ${\rm \AA}^{-1}$.  We define $\left| k_{\delta}\right|$ as a distance between the (0, $\pi$) point and the momentum cut where the measurement has been performed [see Fig. 3(b)].  The reason why we choose this particular cut parallel to the diagonal direction is that $k_{\rm F}$ points of the shadow band and main band are equivalent with respect to the $c$(2$\times$2) zone boundary, (0, $\pi$)-($-\pi$, 2$\pi$) line.  To directly compare the absolute intensity of the shadow band and main band, we plot raw ARPES data after correcting the detector sensitivity.  We did not apply the normalization procedure as used in Fig. 1.  This procedure is reasonable when all data are corrected simultaneously by a wide photoelectron acceptance angle with the manipulator angle fixed with respect to the analyzer.  As seen in Fig. 3(a), we observe an intense steep dispersive main band which clearly crosses $E_{\rm F}$, together with its shadow band counterpart with a suppressed intensity.  These two dispersive bands look symmetric with respect to the folded-zone boundary (white dashed line), confirming the $c$(2$\times$2) nature of the shadow band.  We have fit the MDC at $E_{\rm F}$ with two Lorentzians [Fig. 3(c)] and found that the width of two bands is almost identical to each other while the total spectral weight is about twice different at this momentum region; the estimated ratio of the shadow band to the main band for Pb-Bi2201 (OPT22K) at 25 K is 0.52$\pm$0.06 for $\left| k_{\delta}\right|$ = 0.32 ${\rm \AA}^{-1}$.  To elucidate whether or not the intensity of shadow band is related to important physical parameters, we performed similar analyses for other Pb-Bi2201 samples with different dopings as well as at different temperatures, and also for La-Bi2201.  The results are shown in Fig. 3(d).  To confirm that the estimated weight ratio is not due to the misalignment of the momentum cut in the BZ, we plot data for various $\left| k_{\delta}\right|$ values in Fig. 3(d).  It is noted here that the absolute value of vertical scale does not necessarily have physical importance because of the possible matrix-element effect,\cite{Bansil} but the comparison of the spectral-weight ratio among different experimental conditions would be meaningful since the matrix-element term is not expected to show discernible variation as a function of temperature or doping.  As seen from Fig. 3(d), the spectral-weight ratio shows a similar momentum dependence with a broad maximum of about 60$\%$ at around $\left| k_{\delta}\right|$ = 0.25 ${\rm \AA}^{-1}$ for all the experimental conditions.  This variation of the ratio within a finite $k_{\delta}$ might be due to the $k$ dependence of the matrix element.  It is noticed here that there is no significant doping, temperature, or material dependence of the spectral-weight ratio between the main band and shadow band within an experimental uncertainty of 8$\%$.

\begin{figure}[!t]
\epsfxsize=3.4in
\epsfbox{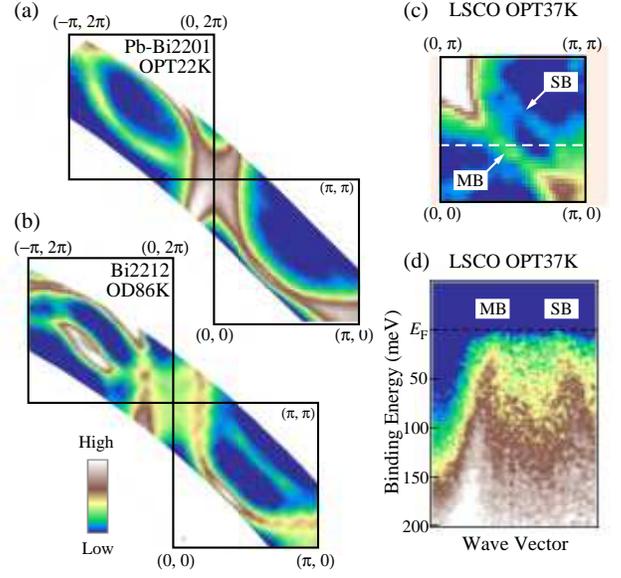}
\vspace{0.25cm}
\caption{
ARPES spectral intensity plots at $E_{\rm F}$ as a function of two-dimensional wave vector for (a) Pb-Bi2201 (OPT22K), (b) slightly overdoped Bi2212 (OD86K), and (c) optimally doped LSCO (OPT37K). (d) ARPES intensity plot of LSCO as a function of binding energy and wave vector measured at a cut shown by a white dashed line in (c).  Temperature during the measurements was kept at 25 K except for Bi2212 (200 K).  In the ARPES measurements we used the He-I$\alpha$ resonance line (21.218 eV) for Bi-based HTSCs and 55.5-eV photons for LSCO.
}
\label{Fig. 4(Color)}
\end{figure}

    To clarify whether the appearance of shadow band is a generic feature of HTSCs, we compare in Figs. 4(a)-4(c) the ARPES intensity profile among different hole-doped HTSCs, (a) Pb-Bi2201 [OPT22K; the same as Fig. 1(b)], (b) Bi2212 (OD86K), and (c) La$_{1.85}$Sr$_{0.15}$CuO$_4$ (OPT37K).  It is clear from Fig. 4(b) that Bi2212 possesses a holelike main band and its shadow band counterpart as Bi2201, although the intensity distribution is much complicated due to the presence of umklapp band.  This experimental result suggests that the origin of shadow band in Bi-based HTSCs is common irrespective of the number of CuO$_2$ planes in a unit cell.  We show in Fig. 4(c) the ARPES intensity at $E_{\rm F}$ of optimally doped LSCO measured with 55.5-eV photons. It is clear from Fig. 4(c) that the optimally doped LSCO also shows a holelike main FS and its shadow band counterpart in the first BZ, although the shadow FS is not so clear as that in Bi-based HTSCs.  To see more clearly the shadow band feature in LSCO, we plot in Fig. 4(d) the ARPES intensity as a function of binding energy and wave vector measured at a cut shown by a white dashed line in Fig. 4(c).  In addition to the main band, we observe another weak band which disperses in the opposite direction compared to the main band, consistent with the band-folding character with respect to the $c$(2$\times$2) zone boundary.  Appearance of the shadow band in LSCO unambiguously demonstrates that the shadow band is not a unique feature of Bi-based HTSCs.  It is noted here that the shadow band in LSCO is not clearly observed at low photon energies including the He-I$\alpha$ resonance line (21.218 eV) unlike the case of Bi2201, but is observed at much higher energy ($h\nu$ = 55.5 eV).   This would be possibly due to the photon-energy dependence of the matrix-element effect reflecting the final-state electronic structure.  We also performed ARPES experiments at $h\nu$ = 55.5 eV with different polarizations of incident light, and found that the experimental result with a linearly polarized light having the polarization vector parallel to the (0, 0)-($\pi$, 0) direction shows a clear signature of the shadow band.

     We now discuss the origin of the shadow band in Bi2201.  The appearance of the shadow band is not explained by the AF picture because of following reasons.  First, the shadow band exists even in a nonsuperconducting heavily overdoped sample where AF correlation is significantly weak.  Second, the intensity of the shadow band is doping-independent while the AF correlation is weakened by hole doping.  Third, the temperature invariance of the shadow band intensity is inconsistent with the general picture that the AF correlation is weaker at higher temperatures.  All these experimental results rule out the AF scenario \cite{refUD} and support the structural origin of the shadow band in Bi-based HTSCs.
	 
	 Now a next question arises as to what structure in the unit cell is responsible for the $c$(2$\times$2) modulation.  Bi2201 possesses single CuO$_2$ layer sandwiched by BiO and SrO layers in a unit cell.  Hence the first candidate is the structural modulation in SrO layers. Magnetic susceptibility experiment in Bi2201 reported that substitution of Sr atoms by rare-earth ($R$) atoms drastically alters the maximum $T_c$ due to existence of cation disorder at Sr site.\cite{Eisaki}  If this structural modulation stabilizes the $c$(2$\times$2) modulation and is responsible for the formation of shadow band in ARPES spectra, the intensity of shadow band would be weaker in samples with higher maximum $T_c$, corresponding to the weaker structural modulation.  Since the maximum $T_c$ of La-Bi2201 is the highest among all $R$-substituted Bi2201 series, it is expected that La-Bi2201 should have a weaker shadow band intensity than Pb-Bi2201.  However, our experimental result does not provide detectable difference of the shadow band intensity between Pb-Bi2201 and La-Bi2201 (Fig. 3), implying that the shadow band is not caused by the structural modulation in SrO layers.  Then the next candidate is a structural modulation in BiO layers.  It is reported by a recent very low-energy electron diffraction (VLEED) experiment that there is faint but finite $c$(2$\times$2) superlattice spots in the VLEED pattern of Bi2212.\cite{Strocov}  Considering the high surface sensitivity of VLEED and ARPES measurements together with the fact that the top-most layer of cleaved surface is a BiO layer, it is possible that the shadow band observed in ARPES originates in the superstructure of the BiO layer at the surface and photoelectrons are diffracted by the $c$(2$\times$2) superstructure.
	 
	 The last possibility is orthorhombic distortion,\cite{Koitzsch} while this distortion would be weak since the hybridization between the shadow band and main band and resultant band bending are not clearly resolved in the present experimental accuracy.  Appearance of the shadow band in optimally doped LSCO [Figs. 4(c) and 4(d)] implies that the final-state diffraction picture associated with the superstructure in BiO layers alone cannot satisfactorily explain the emergence of shadow band, since the BiO layers are absent in LSCO.  We therefore think that, not only the final-state diffraction mechanism, but also the orthorhombic distortion of the crystal (surface) should be taken into account to explain the character of shadow band in Bi2201.  This interpretation is consistent with the x-ray and neutron-diffraction experiments which report that such distortion is indeed present.\cite{LSCOdiffraction}  The above argument is also supported by a recent polarization-dependent ARPES on Bi2212, which reported by the symmetry argument that the shadow band is caused by the orthorhombic distortion.\cite{Mans}  Remaining problems are to clarify whether all the observed shadow band in Bi- and La-based HTSCs are of same structural origin, and to elucidate weather the AF-induced shadow band is present in hole-doped HTSCs.  It is thus strongly desired to elucidate the character of shadow band in LSCO and other class of HTSCs, by performing systematic ARPES experiments covering heavily underdoped region.
				
\section{CONCLUSION}
	We reported systematic high-resolution ARPES results on single-layered Pb- and La-substituted Bi2201 as a function of doping and temperature.  We found that the intensity of shadow band does not alter upon hole doping even when the superconductivity completely disappears.  Observation of the shadow band in La$_{1.85}$Sr$_{0.15}$CuO$_4$ suggests that a simple final-state photoelectron diffraction mechanism associated with the superstructure in BiO layers cannot satisfactorily explain the emergence of shadow band, and the orthorhombic distortion of the crystal should be taken into account.
\\

\acknowledgements
{
This work was supported by JST-CREST and MEXT of Japan. The Synchrotron Radiation Center is supported by NSF DMR Contract No. 9212658.  K.T., S.S., and H.M. thank JSPS for financial support.
}



\begin{references}
\bibitem{refUD}  We note that the possibility of existence of AF-correlation-induced shadow band in underdoped LSCO is not excluded in the present experiment.
\end{references}
\end{document}